
\noindent
{\bf 1. INTRODUCTION}
\bigskip

The detection of gravitational waves is one of the Holy Grails of
modern physics (Schutz 1989; Giazzoto 1989).
Following the pioneering experimental device of J.Weber (see, for example,
Shapiro \& Teukolsky 1983 for a clear introduction and historical references
to Weber's resonant bar detectors)
and many other attempts, a series of new-technology interferometric antennas
is being planned/constructed to achieve higher sensitivities for the
dimensionless wave amplitude {\it h}. These coordinated experimental efforts
 are being complemented by increasingly refined theoretical calculations in
order to study the expected waveforms from each potential source.

Although a number of likely sources has been discussed, there is an underlying
belief (which has, in fact, strongly influenced the design of the detectors)
that non-spherical supernovae and coalescing binaries
are among the strongest candidates for objects that the devices
should detect. The purpose of this work is to argue quantitatively that
wobbling neutron stars, to which comparatively little attention has been paid,
may be an equally good bet regarding the probability of detection.
\bigskip
\noindent
{\bf 2. DISCUSSION AND CALCULATION}
\bigskip

Rotating neutron stars will emit gravitational waves by means of a
time-dependent quadrupole moment, generated either by the lack of body
symmetry on the equatorial plane or by precession caused by a misalignment of
the spin and symmetry axes. It is presently quite uncertain how asymmetric
a pulsar can be and we shall not refer to the former case (but keep it in
mind as a true possibility) in this work. The latter wobbling neutron stars
will emit mainly at frequencies close to the rotational one $f$ if the
wobble angle $\theta$ is small.
The resulting amplitude of the waves is calculated as
$h \; = \; (16 \, \pi \, G \, F / c^{3} \, \Omega^{2})^{1/2}$ ; where
the flux is $F \, = \, L_{GW} /4 \pi \, r^{2}$ and $\Omega = 2 \pi \, f$
is the angular frequency of the pulsar (Zimmermann 1978) yielding

$$ h = 1.4 \; \times \; 10^{-18} \; \epsilon \; \theta \;
\bigl({{I_{zz}} \over {10^{45} \; g \; cm^{2}}}\bigr) \;
f_{kHz}^{2} \; r_{kpc}^{-1} \eqno (1) $$

where

$$ \epsilon = (I_{zz} \; - \; I_{xx})/ I_{zz}  $$

$\theta$ is the wobble angle, $f_{kHz}$ is the frequency in kHz,
$r_{kpc}$ is the distance to the star in kpc, $I_{zz}$ is the moment of inertia
with respect to the rotation axis and $I_{xx}$ is any of the moments of inertia
orthogonal to it. This equation has been established using the "slow-motion"
approximation for the gravitational energy output of the star (Misner, Thorne
\& Wheeler 1973) but is
probably accurate up to a factor $\sim \; 2$ in the rapidly rotating regime
(Zhong 1985).
For a given pulsar, distance and structure, the wobble angle can be used to
parametrize the expected amplitude of eq.(1).
According to Pines \& Shaham (1974) and Zimmermann (1978)
an upper physical limit to $\theta$ is
$\theta_{M} \; \sim \; 10^{-1}$, and values of
$\theta \; \sim \; 10^{-2} - 10^{-3}$ may be considered as moderate.

The expected sensitivity of the interferometric detectors to the waves
(Vogt 1989; Giazzoto 1991)
is of the order of $h \; \sim \; 10^{-22}$
for short-lived, impulsive bursts ; and  $h \; \sim \; 10^{-25} - 10^{-26}$
for periodic sources in which integration times of about $10^{7} \; s$ are
possible. According to eq.(1), the emission expected from a pulsar
undergoing a wobble motion characterized by $\theta$ needs also an
accurate estimate of the quantity $\epsilon$.

To study the strenght of the expected emission we have performed,
as discussed in de Ara\'ujo et al. (1993, hereafter paper I),
fully relativistic calculations of the stellar structure base on the approach
of Butterworth \& Ipser (1976) and Butterworth (1976)
(see also Friedman, Ipser \& Parker 1986).
Our results indicate that the values of the
gravitational ellipticity $\epsilon$ due to the stellar rotation
are typically one or two orders of magnitude higher
than those usually {\it adopted} in
the literature, the latter being appropriate for slower radio pulsars
(remarkably, comparable results are already implicit,
for example, in Friedman, Ipser \& Parker 1986). As shown below, this feature
leads to outputs of {\it h} which may
be detectable by the upcoming gravitational antennas generation if those
pulsars deformed by rotation can be induced to precess by external/internal
torques.

Table 1 gives the stellar parameters calculated at fixed baryon number
(that corresponding to $M \; = \; 1.4 \; M_{\odot}$ for a static star)
for different rotation velocities using the
medium-stiff Bethe-Johnson I equation of state (Bethe \& Johnson 1974)
for the neutron matter.
While it is not guaranteed that the actual composition of the pulsars can
be represented by this choice, it is generally agreed that the chosen
equation of state is a reasonable compromise given the present uncertainties
on the subject and it will be employed for the sake of definiteness.

In order to relate the putative gravitational wave emission to the expected
signal at the detectors it is imperative to estimate the damping timescale
for a given wobbling pulsar due to the emission of gravitational waves.
The characteristic braking timescale is

$$ \tau_{brake} \; \simeq \;
{5 \; c^{5} \over {128 \; G \; I_{zz} \; {\Omega}_{o}^{4} \; {\epsilon}^{2}
\; }} \simeq 2 \, \bigl( {\epsilon \over {0.1}} \bigr)^{-2} \;
\bigl( {I_{zz} \over {10^{45} \, g \, cm^{2}}} \bigr)^{-1} \;
\bigl( {\Omega_{o} \over {5000 \, s^{-1}}}\bigr)^{-4} \, s  \eqno (2) $$

where $\Omega_{o}$ is the initial rotational angular velocity.
Therefore, even though the sources would have an explicit periodic behaviour,
the fact that $\tau_{brake} \, \ll \, 10^{7} \, s$ qualifies them as
"impulsive" or burst one since the duration of the emission is short
compared to the experimental observation time (see a full discussion in
Thorne 1987). If we require $h \, \geq \, 10^{-22}$ (the condition of
burst detectability foreseen for LIGO-type interferometers)
for a $P \, = \, 2 ms$
pulsar with a "fiducial" value of  $I_{zz} = 10^{45} \, g \, cm^{2}$ we
would need

$$ {\theta \over {r_{kpc}}} \, \geq \, 6 \, \times \, 10^{-3} \eqno(3) $$

Thus, observation of a "spike" burst with duration limited by $\tau_{brake}$
from anywhere in the Galaxy ($r \, \simeq \, 20 \, kpc$) would require
wobble angles of the order of the more extreme theoretical expectations.
However, it is now known that the population of $ms$ pulsars is large
(see e.g. Kulkarni \& Thorsett 1993) and there have been (unexpectedly)
observed in nearby clusters like 47 Tuc (see e.g. Lyne 1992)
which is only $\sim \, 4 \, kpc$ away.
Those potential sources would require values of $\theta \, \simeq \, 10^{-2}$
to be detected at LIGO-type observatories. We remark that a reliable
calculation of $\epsilon$ is an important ingredient for such a conclusion.
In addition, not only the interesting wobble angles
are probably less than extreme, but also that the
emission will come out at frequencies
which are {\it not} expected (in principle)
to suffer from severe noise problems
in contrast with the values required to detect the precession of slower
Crab-like sources.

It should be stressed that the actual number of sources in
a sphere of radius $20 \; kpc$ (roughly the volume of the Galaxy) will depend
on the details of the internal dynamics and external torques and can not be
reliable evaluated other than in a statistical way. The total number of
pulsars in the sample volume may be as high as
$10^{8}$ and it has been estimated (Kulkarni, Narayan \& Romani 1990)
that the subpopulation of $ms$ pulsars in globular clusters only is higher
than $10^{4}$. Detailed calculations of external (i.e. encounters with
perturbing stars) and internal (i.e. phase transitions) mechanisms capable
of exciting moderate wobble motions must be undertaken to address the expected
number of sources at a given time. As an example, mechanisms which may
conceivably produce small wobble motions in the range
$\theta \, \sim \, 10^{-4} - 10^{-6}$ have been discussed in Pines \& Shaham
(1974).
Several other excitation mechanisms
may be operative as well.
If Nature provides even a single source out of the whole galactic
population (the case for a precessing Her X-1 and newer data
from radio pulsars seem to suggest the
possibilty of wobble being an ubiquitous phenomenon)
it may be enough greatly to improve our knowledge on gravitational
wave phenomena.

\vskip 1 true cm
\noindent
{\bf Acknowledgements}

We would like to acknowledge the financial support of the Funda\c c\~ao
de Amparo \`a Pesquisa do Estado de S\~ao Paulo (FAPESP) and the Conselho
Nacional de Desenvolvimento Cient\'\i fico e Tecnol\'ogico (CNPq). The kind
hospitality given to M.C., J.A.F.P. and J.E.H.
during scientific visits to the VIRGO
project at INFN, Pisa (Italy)
is also gratefully acknowledged. J.C.N.A. would like
to thank the hospitality received in the SISSA/ISAS, in particular from
the Head of the Astrophysical Sector, Prof. D.W. Sciama.
M.C. acknowledges
the financial support of the CAPES (Brazil).
We also thank our referee, Dr. B.Schutz, for correcting a mistake of the
damping time-scale made on the first version and a careful reading of the
paper.
Finally we would
like to thank Dr.W.Velloso for his encouragement and advice.
\vfill\eject

Table 1. Stellar parameters of rotating B-J I pulsars

$$ \vbox {\settabs 5\columns

\+ $\Omega \; (rad \; s^{-1})$ &   e   & $\epsilon$  &
$ I_{zz} \; (10^{45} \; g \; cm^{2})$
& $ I_{xx} \; (10^{45} \; g \; cm^{2}) $ \cr
\+ 3000   &   0.31  &  0.047  &  0.6557  &  0.6246 \cr
\+ 4030   &   0.38  &  0.089  &  0.6994  &  0.6374 \cr
\+ 5000   &   0.51  &  0.14   &  0.7763  &  0.6658 \cr
\+ 6203   &   0.75  &  0.25   &  0.9946  &  0.7476 \cr} $$
\vskip 2 true cm
\noindent
{\bf Table captions}

\vskip 1 true cm
\noindent
{\bf Table 1.} The relevant parameters for gravitational waves emission from
wobbling pulsars. Here e is the usual stellar eccentricity and the other
quantities are defined in the text. Note that the maximum value of $\Omega$
corresponds to the Keplerian value although several instabilities may limit
the actual maximum rotation rate to about $0.9$ of the former.
\vfill\eject
\noindent
{\bf References}

\bigskip
\noindent
Bethe,H.A. , Johnson,M., 1974, Nuc.Phys.A, 230, 1.
\noindent
Butterworth E.M., 1976, ApJ, 204, 561.

\noindent
Butterworth E.M., Ipser J.R., 1976, ApJ, 204, 200.

\noindent
de Ara\'ujo J.C.N., de Freitas Pacheco J.A., Cattani M., Horvath J.E., 1993,
submitted.

\noindent
Friedman J.L., Ipser J.R., Parker L., 1986, ApJ 304, 105.

\noindent
Giazzoto A., 1989, Phys. Rep., 182, 367

\noindent
Kulkarni S., Narayan R., Romani R., 1990, ApJ, 356, 174.

\noindent
Kulkarni S., Thorsett S., 1993, Nat, 364, 579

\noindent
Lyne A.G., 1992, in X-ray Binaries and Recycled Pulsars. NATO:
ASI Series (eds van den Heuvel, E.P.J. \& Rappaport, S.A.), pg. 79
(Kluwer: Dordrecht)

\noindent
Misner C.W., Thorn K.S., Wheeler J.A., 1973, Gravitation, W.H.Freeman \& Co.,
San Francisco.

\noindent
Pines D., Shaham J., 1974, Comm. Ap., 2, 37.

\noindent
Schutz B., ed., 1989, The Detection of Gravitational Waves Data Analysis,
Kluwer, Dordretch.

\noindent
Shapiro S., Teukolsky S.L., 1983, Black Holes, White Dwarfs and Neutron Stars:
 the Physics of Compact Objects, J.Wiley \& Sons, New York.

\noindent
Thorne K.S., 1987, in Hawking S., Israel W., eds., 300 Years of Gravitation,
Cambridge University Press, London.

\noindent
Vogt R., 1989, talk given at the 12th International Conference on General
Relativity and Gravitation (GR12).

\noindent
Zimmermann M., 1978, Nat, 271, 524.

\noindent
Zhong M-Q., 1985, Acta Ap Sin, 5, 103.
\bye